\documentclass[aps,showpacs,twocolumn,groupedaddress]{revtex4}
\usepackage{graphics}
\usepackage{epsfig}
\usepackage{epsf}
\usepackage{amssymb}
\usepackage{epstopdf}
\usepackage{amsmath}
\usepackage{url}

\input{tcilatex}
\begin{document}

\title{Comment on \textquotedblleft Spatial Coherence and Optical Beam
Shifts\textquotedblright }
\author{Li-Gang Wang$^{1}$, Shi-Yao Zhu$^{2}$, and M. Suhail Zubairy$^{3}$}
\affiliation{$^{1}$Department of Physics, Zhejiang University, Hangzhou 310027, China\\
$^{2}$Beijing Computational Science Research Center, Beijing, 100084, China\\
$^{3}$Institute for Quantum Science and Engineering (IQSE) and Department of
Physics and Astronomy, Texas A$\&$M University, College Station, TX
77843-4242, USA }
\date{\today }
\maketitle

A recent letter by L\"{o}ffler \textit{et al.} \cite{Loffler2012} presents
experimental results concluding that the degree of spatial coherence does
not influence the spatial Goos-H\"{a}nchen (GH) shift. The authors measure
the difference of the GH shift, $D_{p}-D_{s}$, for $p$- and $s$-polarization
of partially coherent light, but the absolute GH shift, $D_{p}$ or $D_{s}$,
is not presented. However in our papers \cite{WANGLG2008,WANGLG2012}, we
show the absolute GH shift, $D_{p}$ or $D_{s}$, is affected by spatial
coherence. Their experimental result shows that the measured value, $%
D_{p}-D_{s}$, is independent of the ratio $\sigma _{g}/\sigma _{s}$, where $%
\sigma _{g}$ and $\sigma _{s}$ are the spatial coherence and the beam width,
respectively. They then conclude that "this demonstrates that the
theoretical result in Refs. \cite{Simon1989,Aiello2011,Aiello2012} is
correct, contrary to competing claims \cite{WANGLG2008,WANGLG2012}" and
hence the "dispute in the literature \cite%
{Simon1989,Aiello2011,Aiello2012,WANGLG2008,WANGLG2012} is now definitively
resolved". We disagree.

This comment is to show that our simulation data, based on the theory and
method in Ref. \cite{WANGLG2008}, are also in agreement with their
experimental data presented for $D_{p}-D_{s}$ in \cite{Loffler2012}. Then we
also demonstrate how to show the effect of spatial coherence on the GH
shifts.

Before presenting our results, we point out that the incident partially
coherent fields, Eq. (7) in Ref. \cite{WANGLG2008}, can be well approximated
from Eq. (3) in Ref. \cite{Aiello2012} due to two facts: (a) the
dimensionless quantities, $y_{1,2}(\sin \theta )/(k\alpha )$, are extremely
small in laboratory coordinates, so that we have $\gamma _{1,2}=\alpha
-y_{1,2}\sin \theta /(2k)=\alpha \lbrack 1-y_{1,2}\sin \theta /(2k\alpha
)]\rightarrow \alpha $ for Eq. (4) in Ref. \cite{Aiello2012}; (b) the
higher-order phase term, exp$\left\{ \frac{ik\cos ^{2}\theta \sin \theta
y_{1}y_{2}(y_{2}-y_{1})}{8k^{2}(\alpha ^{2}-\beta ^{2})}\right\} $, can be
replaced by unity in the presence of the term exp$\left\{ -ik\sin \theta
(y_{2}-y_{1})\right\} $. Furthermore, Eq. (7) in Ref. \cite{WANGLG2008}
describes the field distribution at the interface, which can be assumed to
be of any shape including Gaussian as well as a point-like source. Our
method is exact and is valid in both paraxial and non-paraxial regimes (see
Eq. (4) in Ref. \cite{WANGLG2008}).

\begin{figure}[tbp]
\includegraphics[width=8cm]{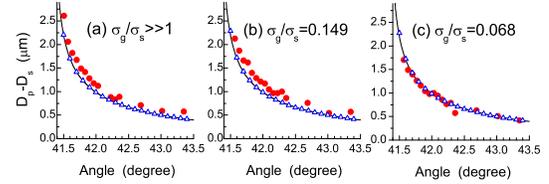}
\caption{(Color online) Comparision of our simulation data and their
prediction with the experimental data \protect\cite{Loffler2012}. Here solid
dots are experimental data, open triangles are our simulation data, and
solid curves are the predictions of Refs. \protect\cite%
{Simon1989,Aiello2011,Aiello2012}. }
\label{Fig:fig1}
\end{figure}

In Fig. 1, we compare our simulation data with their experimental results 
\cite{Loffler2012} and their prediction curve for the three cases: (a) $%
\sigma _{g}/\sigma _{s}\gg 1$ ($\sigma _{s}\approx 0.4$mm), (b) $\sigma
_{g}/\sigma _{s}=0.149$ ($\sigma _{s}\approx 0.9$mm), and (c) $\sigma
_{g}/\sigma _{s}=0.068$ ($\sigma _{s}\approx 2.1$mm). It is clear from Fig.
1, that our simulation results are nearly the same as their theoretical
results, and both have only small difference from the experimental data.
Therefore their experimental data should not form the basis of an objection
to our claims in \cite{WANGLG2008,WANGLG2012}.

\begin{figure}[tbp]
\includegraphics[width=7cm]{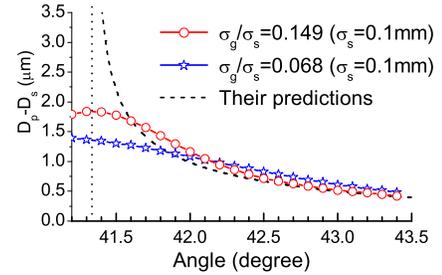}
\caption{(Color online) Effect of spatial coherence on $D_{p}-D_{s}$.}
\label{Fig:fig2}
\end{figure}

A question of interest is: How can we obtain a substantial difference
between our simulation and their predictions? We note that, in the
experiments discussed above, although $\sigma _{g}/\sigma _{s}$ is
considerably small, the absolute values, $\sigma _{g}$ and $\sigma _{s}$,
are much larger than the wavelength, $\lambda $. If $\sigma _{g}$ is close
to $\lambda $ but is still larger than $\lambda $, and $\sigma _{g}/\sigma
_{s}$ is fixed, then $D_{p}$, $D_{s}$, and even $D_{p}-D_{s}$ in our
simulation and their prediction will be significantly different especially
near the critical angle as shown in Fig. 2. The degree of spatial coherence $%
\sigma _{g}$ will then have a large effect on the GH shift. This is
basically our claim based on the theory presented in Refs. \cite%
{WANGLG2008,WANGLG2012}.

\begin{acknowledgments}
This work is supported the National Basic Research Program of China (No.
2012CB921602) and by NSFC grants (No. 61078021, 11174026 and No. 11274275).
The research of MSZ is supported by NPRP grant 4-346-1-061 by the Qatar
National Research Fund (QNRF).
\end{acknowledgments}


\begin{thebibliography}{9}
\bibitem{Loffler2012} W. L\"{o}ffler, A. Aiello, and J. P. Woerdman, Phys.
Rev. Lett. \textbf{109}, 213901 (2012).

\bibitem{WANGLG2008} L.-Q. Wang, L.-G. Wang, S.-Y. Zhu, and M. S. Zubairy,
J. Phys. B \textbf{41}, 055401 (2008).

\bibitem{WANGLG2012} L.-G. Wang and K.-H. Liu, Opt. Lett. \textbf{37}, 1056
(2012).

\bibitem{Simon1989} R. Simon and T. Tamir, J. Opt. Soc. Am. A \textbf{6}, 18
(1989).

\bibitem{Aiello2011} A. Aiello and J. P. Woerdman, Opt. Lett. \textbf{36},
3151 (2011).

\bibitem{Aiello2012} A. Aiello and J. P. Woerdman, Opt. Lett. \textbf{37},
1057 (2012).
\end{thebibliography}
\end{document}